\documentclass[twoside,twocolumn,9pt]{article}
\usepackage{extsizes}
\usepackage[super,sort&compress,comma]{natbib} 
\usepackage[version=3]{mhchem}
\usepackage[left=1.5cm, right=1.5cm, top=1.785cm, bottom=2.0cm]{geometry}
\usepackage{balance}
\usepackage{mathptmx}
\usepackage{sectsty}
\usepackage{graphicx} 
\usepackage{lastpage}
\usepackage[format=plain,justification=justified,singlelinecheck=false,font={stretch=1.125,small,sf},labelfont=bf,labelsep=space]{caption}
\usepackage{float}
\usepackage{fancyhdr}
\usepackage{fnpos}
\usepackage[english]{babel}
\addto{\captionsenglish}{%
  
}
\usepackage{array}
\usepackage{droidsans}
\usepackage{charter}
\usepackage[T1]{fontenc}
\usepackage[usenames,dvipsnames]{xcolor}
\usepackage{setspace}
\usepackage[compact]{titlesec}
\usepackage{hyperref}
\usepackage{mathrsfs}

\usepackage{epstopdf}

\definecolor{cream}{RGB}{222,217,201}

\begin{document}

\pagestyle{fancy}
\thispagestyle{plain}
\fancypagestyle{plain}{
\renewcommand{\headrulewidth}{0pt}
}

\makeFNbottom
\makeatletter
\renewcommand\LARGE{\@setfontsize\LARGE{15pt}{17}}
\renewcommand\Large{\@setfontsize\Large{12pt}{14}}
\renewcommand\large{\@setfontsize\large{10pt}{12}}
\renewcommand\footnotesize{\@setfontsize\footnotesize{7pt}{10}}
\makeatother

\renewcommand{\thefootnote}{\fnsymbol{footnote}}
\renewcommand\footnoterule{\vspace*{1pt}%
\color{cream}\hrule width 3.5in height 0.4pt \color{black}\vspace*{5pt}} 
\setcounter{secnumdepth}{5}

\makeatletter 
\renewcommand\@biblabel[1]{#1}            
\renewcommand\@makefntext[1]%
{\noindent\makebox[0pt][r]{\@thefnmark\,}#1}
\makeatother 
\renewcommand{\figurename}{\small{Fig.}~}
\sectionfont{\sffamily\Large}
\subsectionfont{\normalsize}
\subsubsectionfont{\bf}
\setstretch{1.125} 
\setlength{\skip\footins}{0.8cm}
\setlength{\footnotesep}{0.25cm}
\setlength{\jot}{10pt}
\titlespacing*{\section}{0pt}{4pt}{4pt}
\titlespacing*{\subsection}{0pt}{15pt}{1pt}

\fancyfoot{}
\fancyfoot[LO,RE]{\vspace{-7.1pt}\includegraphics[height=9pt]{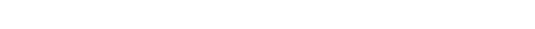}}
\fancyfoot[CO]{\vspace{-7.1pt}\hspace{11.9cm}\includegraphics{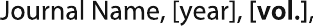}}
\fancyfoot[CE]{\vspace{-7.2pt}\hspace{-13.2cm}\includegraphics{head_foot/RF}}
\fancyfoot[RO]{\footnotesize{\sffamily{1--\pageref{LastPage} ~\textbar  \hspace{2pt}\thepage}}}
\fancyfoot[LE]{\footnotesize{\sffamily{\thepage~\textbar\hspace{4.65cm} 1--\pageref{LastPage}}}}
\fancyhead{}
\renewcommand{\headrulewidth}{0pt} 
\renewcommand{\footrulewidth}{0pt}
\setlength{\arrayrulewidth}{1pt}
\setlength{\columnsep}{6.5mm}
\setlength\bibsep{1pt}

\makeatletter 
\newlength{\figrulesep} 
\setlength{\figrulesep}{0.5\textfloatsep} 

\newcommand{\topfigrule}{\vspace*{-1pt}%
\noindent{\color{cream}\rule[-\figrulesep]{\columnwidth}{1.5pt}} }

\newcommand{\botfigrule}{\vspace*{-2pt}%
\noindent{\color{cream}\rule[\figrulesep]{\columnwidth}{1.5pt}} }

\newcommand{\dblfigrule}{\vspace*{-1pt}%
\noindent{\color{cream}\rule[-\figrulesep]{\textwidth}{1.5pt}} }

\makeatother

\twocolumn[
  \begin{@twocolumnfalse}
{\includegraphics[height=30pt]{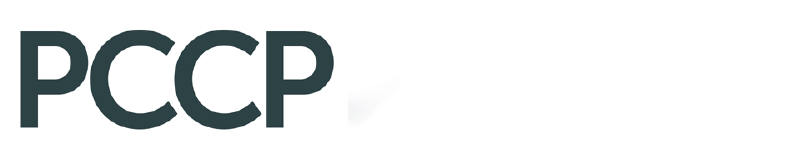}\hfill\raisebox{0pt}[0pt][0pt]{\includegraphics[height=55pt]{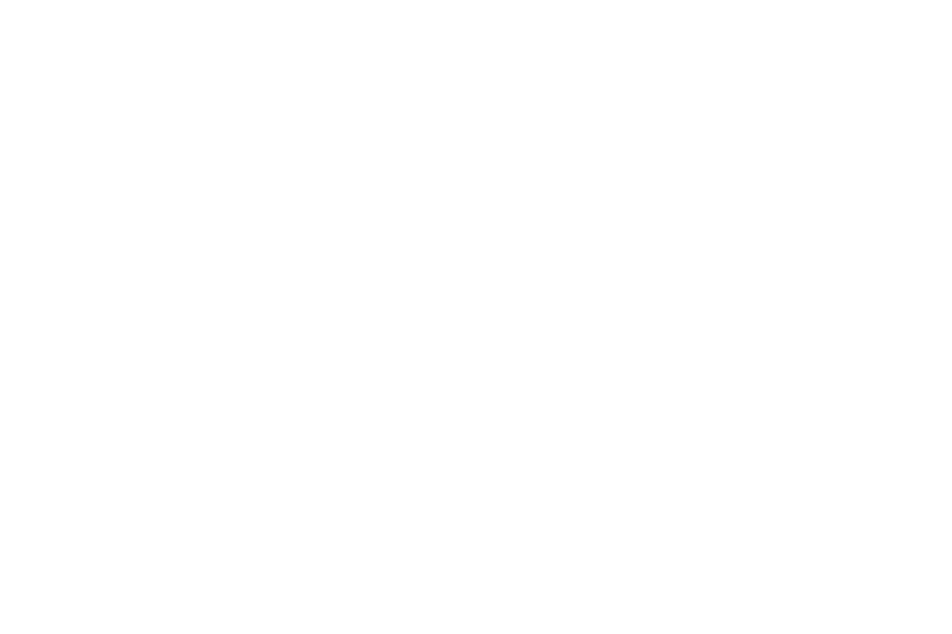}}\\[1ex]
\includegraphics[width=18.5cm]{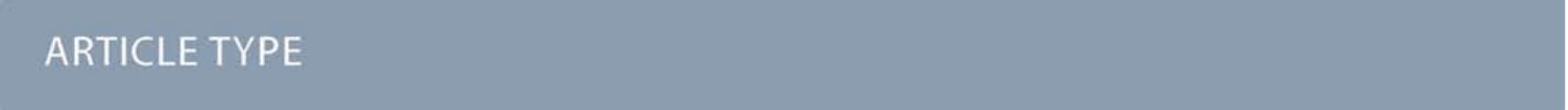}}\par
\vspace{1em}
\sffamily
\begin{tabular}{m{4.5cm} p{13.5cm} }

\includegraphics{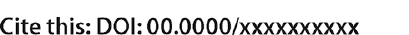} & \noindent\LARGE{\textbf{Data-driven analysis of the electronic-structure factors controlling the work functions of perovskite oxides}} \\
\vspace{0.3cm} & \vspace{0.3cm} \\

 & \noindent\large{Yihuang Xiong,\textit{$^{a}$} Weinan Chen,\textit{$^{a}$} Wenbo Guo,\textit{$^{b}$} Hua Wei,\textit{$^{b}$} Ismaila Dabo\textit{$^{a,c}$}} \\

\includegraphics{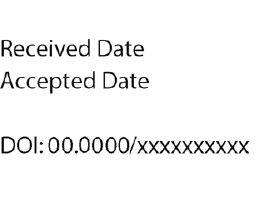} & \noindent\normalsize{Tuning the work functions of materials is of practical interest for maximizing the performance of microelectronic and (photo)electrochemical devices, as the efficiency of these systems depends on the ability to control electronic levels at surfaces and across interfaces.  Perovskites are promising compounds to achieve such control. In this work, we examine the work functions of more than 1,000 perovskite oxide surfaces (ABO$_3$) by data-driven (machine-learning) analysis and identify the factors that determine their magnitude. While the work functions of BO$_2$-terminated surfaces are sensitive to the energy of the hybridized oxygen $p$ bands, the work functions of AO-terminated surfaces exhibit a much less trivial dependence with respect to the filling of the $d$ bands of the B-site atom and of its electronic affinity. This study shows the utility of interpretable data-driven models in analyzing the work functions of cubic perovskites from a limited number of electronic-structure descriptors.} \\

\end{tabular}

 \end{@twocolumnfalse} \vspace{0.6cm}

  ]

\renewcommand*\rmdefault{bch}\normalfont\upshape
\rmfamily
\section*{}
\vspace{-1cm}


\footnotetext{\textit{$^{a}$~Department of Materials Science and Engineering, and Materials Research Institute,
The Pennsylvania State University, University Park, PA, USA, E-mail: YihuangXiong@psu.edu}}
\footnotetext{\textit{$^{b}$~College of Information Sciences and Technology, The Pennsylvania State University, University Park, PA 16802, USA }}
\footnotetext{\textit{$^{c}$~Institutes of Energy and Environment, The Pennsylvania State University, University Park, PA 16802, USA }}

\footnotetext{\dag~Electronic Supplementary Information (ESI) available: A schematic of a regression tree in the random forests, feature selections based on correlation matrix and importance ranking, and projected density of states of selected perovskites. See DOI: 10.1039/cXCP00000x/}


\section{Introduction}
\label{sec:Introduction}
\begin{figure}
 	\centering
 	\includegraphics[width=0.48\textwidth]{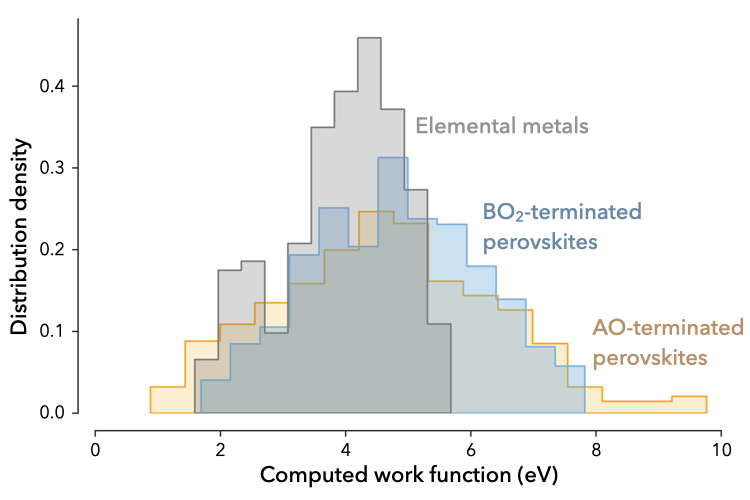}    
 	\caption{Distribution of the computed work functions for elemental metals~\cite{TRAN201948} and cubic perovskites. Perovskites show a broad distribution of work functions.}
 	\label{WF_dist_plot}
\end{figure}

The work function measures the energy of extracting an electron from a material. Understanding trends in the work function is technologically important to thermionics~\cite{Yamamoto_2005}, optoelectronics, electrochemistry, and photocatalysis~\cite{Castelli2014,RobustCO2,PRM_XiongDabo,TRASATTI1972163,C7CP03081A,Songe1700336,Shao_Horn751}---with one primary example being the possibility to optimize the activity of a surface by tuning its electronic affinity \cite{tune_WF_electrode}. Perovskites are a remarkably versatile class of materials that can be synthesized with controlled purity and relatively high yield~\cite{ray_toolbox,SrVO3,doi:10.1002/adfm.201602767}. Due to the interplay between their structural, chemical, and electronic characteristics, perovskites are promising candidates for achieving sensitive control of the work function. Figure \ref{WF_dist_plot} compares the work functions of elemental metals \cite{TRAN201948} with those of perovskite oxides; it is apparent that perovskites show a wide distribution of work functions, providing a rich compositional space for the design of {\it e.g.} thermionic converters (requiring low work functions)~\cite{dane_WF} and photovoltaic hole collectors (requiring high work functions). 

\begin{figure}
 	\centering
 	\includegraphics[width=0.48\textwidth]{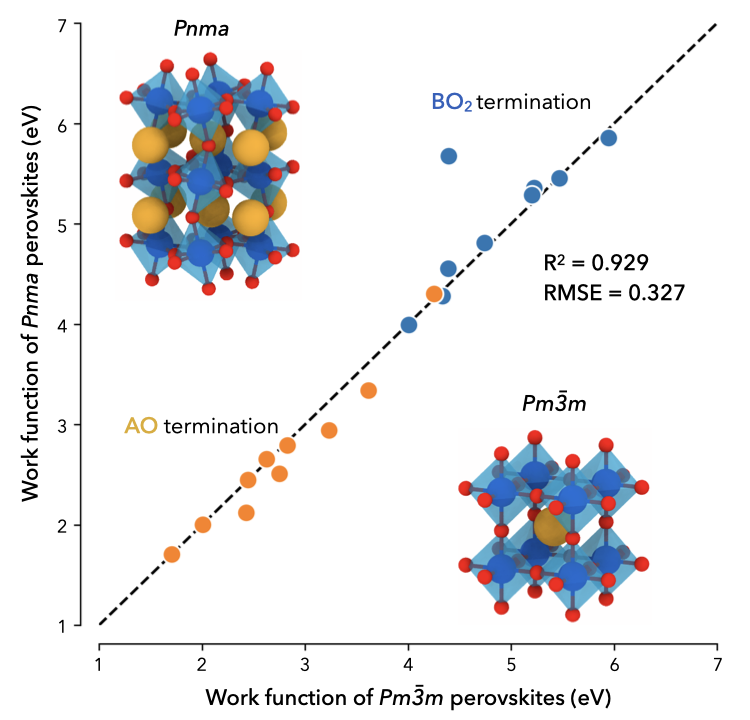}    
 	\caption{Comparison of computed work functions along the [001] direction for a representative set of 10 perovskite oxides with space groups $Pm\bar{3}m$ and $Pnma$. The selected compositions are ABO$_3$ where A  = Ca or Sr and B = Ti, V, Cr, Mn or Fe. The overall coefficient of determination (R$^{2}$) and root mean squared error are of 0.929 and 0.237 eV, respectively.}
 	\label{WF_comparison}
\end{figure}

In this work, we develop a data-driven understanding of the work functions of perovskite oxides in their prototypical cubic symmetry. For comparison, Fig.~\ref{WF_comparison} shows the work functions of 10 representative perovskites in the orthorhombic ($Pnma$) and cubic ($Pm\bar{3}m$) phases along their [001] surface facets. These results highlight a strong correlation between the work functions of these structures, indicating that the high-symmetry, cubic phase may provide a reliable basis to infer the work functions of low-symmetry, perovskite-related structures featuring octahedral rotations. Examining cubic structures is also relevant to high-entropy perovskites~\cite{Curtarolo} that tend to spontaneously adopt high symmetry~\cite{JIANG2018116, Curtarolo}. We thus present a detailed analysis of the dependence of the work functions of cubic perovskites as a function of composition and termination using extensive computational data.

\section{Computational method}
\label{sec:computational_methods}
\subsection{Crystal structures and first-principles calculations}
\label{sec:  Crystal structures and first-principles simulations}

\begin{figure}
 	\centering
 	\includegraphics[width=0.48\textwidth]{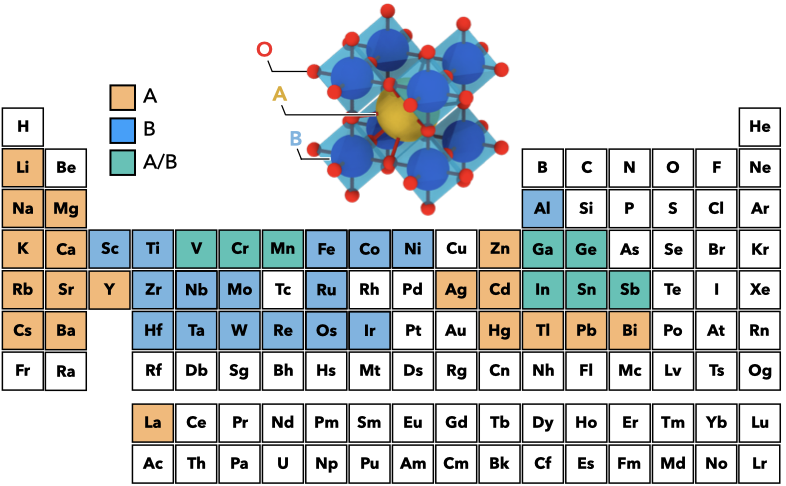}    
 	\caption{ A perovskite unit cell is composed of A cation in the center of the unit cell, and B cation octahedrally coordinated with the oxygen (top). The elemental compositions that are used to construct the perovskites are highlighted in the periodic table (bottom).}
 	\label{Perovskite_comp}
\end{figure}

A perovskite crystal structure with formula ABO$_{3}$ is shown in Fig.~\ref{Perovskite_comp}. The B-site cation is octahedrally coordinated to oxygen, and, typically, a larger A-site cation adopts a twelve-fold coordination with the surrounding oxygen atoms. The cubic perovskite phase exists in nature ({\it e.g.} SrTiO$_3$ and SrVO$_3$), while many other lower-symmetry stable structures are also found. Compared to the ideal cubic structure, distortions such as octahedral rotations and cations displacements may occur, and some of them are responsible for functional properties such as ferroelectricity~\cite{no_ferroelectricity}. Nevertheless, we adopt the cubic phase as a simple template for statistical analysis, as discussed above and justified in Fig.~\ref{WF_comparison}.

\begin{figure}
 	\centering
 	\includegraphics[width=0.48\textwidth]{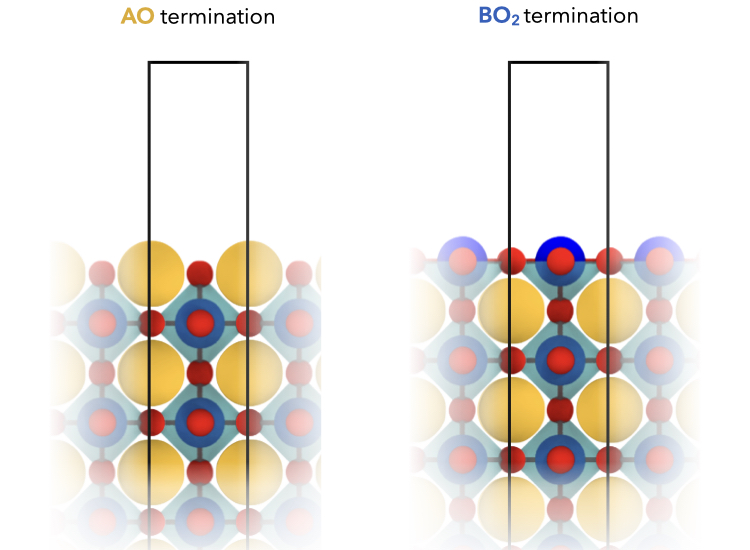}    
 	\caption{Surface structures considered in this work: the AO and BO$_2$ terminations.}
 	\label{A_B_interface}
\end{figure}

Following Refs.~\citenum{Castelli_probability} and \citenum{Wolverton_HT}, we select the constituent metal cations based on their propensity to form a stable cubic phase. The elements that are considered in this work are highlighted in Fig.~\ref{Perovskite_comp}. The A-site elements include the main-group metals, while the majority of the B-site elements belong to the transition metals series. Considering the alternating AO and BO$_2$ layers, we construct two types of interfaces along the [001] direction, as shown in Fig.~\ref{A_B_interface}. Using the optimized bulk structures, each slab geometry is built symmetrically with 9 ionic layers. The periodic slabs are separated by 14 \AA\ of vacuum. Only the two outermost layers are allowed to move during geometry optimization.

All first-principles calculations are managed by the AiiDA high-throughput calculation infrastructure~\cite{pizzi2016aiida}. The self-consistent-field calculations are performed at the semilocal Perdew--Burke--Ernzerhorf (PBE) level~\cite{perdew1996generalized} using the {\sc pw} code of the Q{\sc uantum} ESPRESSO distribution~\cite{giannozzi2009quantum}. Ionic cores are represented by norm-conserving pseudopotentials with kinetic-energy cutoffs of 100 Ry for the reciprocal-space expansion of the wave functions~\cite{rappe1990optimized}. Bulk structures are fully optimized through variable-cell optimization, while sampling the Brillouin zone with a $\Gamma$-centered Monkhorst--pack grid of 12$\times$12$\times$12~\cite{monkhorst1976special}. For slab calculations, a Marzari-Vanderbilt cold smearing of 0.01 Ry~\cite{marzari1999thermal} is employed to discretize the Brillouin zone with a reduced ${\bf k}$-points mesh of 6$\times$6$\times$1. In addition, the {\sc Environ} module is applied to automatically align the Fermi level with respect to vacuum~\cite{dipole_correction, Dabo_Kozinsky,Dabo_Li}. The atomic positions are then fully optimized until the interatomic forces are smaller than 0.02 eV/\AA. 

Based on the optimized perovskite surface, we can calculate the work functions as
\begin{equation}
    \Phi = \Phi^\circ-{E_{\rm F}},
\end{equation}
where  $\Phi^\circ$ is the potential in vacuum and $E_{\rm F}$ is the Fermi energy.

Due to the semilocal PBE approximation, the calculated band gap and work function are expected to be underestimated~\cite{Singh-Miller,TRAN201948}. Even though previous work suggests that the work functions of metals calculated from PBE are consistent with experimental measurements~\cite{TRAN201948,Stroppa_2008}, there is still debate about the accuracy of PBE work functions for perovskite oxides. More generally, the work functions of metal oxides can be strongly influenced by surface orientations, terminations, and defects. Studies by Ma~\textit{et al.} and Guo~\textit{et al.} showed that an accurate description of band gaps could lead to improved predictions of the work functions and band edges of semiconductors~\cite{Guo_band_alignment,Ma_STO}. Although predicting absolute work functions using the PBE functional may not be accurate, Ma~\textit{et al.}, and Chambers and Sushko showed that the PBE approximation is reliable to estimate differences in the work functions of AO and BO${_2}$-terminated surfaces~\cite{Ma_STO,Chambers_STO}. While beyond the scope of this work, it is expected that hybrid functionals such as Heyd-Scuseria-Ernzerhof (HSE)~\cite{hse} could be more accurate for evaluating the work functions of perovskite metal oxides. Since the goal of this study is to understand trends between work functions and electronic descriptors,  we argue that it is suitable to use the PBE functional.
 
\subsection{Machine-learning method and descriptor selection}
\label{sec: Machine learning method}

On the basis of our computed datasets, we aim to identify the features that best describe the work functions of the perovskites. To achieve this, we employ a statistical learning method. We chose our model based on interpretability and performance. Here we use random forest regression~\cite{liaw2002classification}, which is an ensemble statistical learning method that integrates a number of decision trees and that returns the average prediction of these trees~\cite{safavian1991survey}. In specific terms, given a training set ($\mathbf{X}$, $\mathbf{y}$) where $\mathbf{X}$ is the features and $\mathbf{y}$ are the corresponding responses, the random forest model is trained by repeatedly sampling a subset $\mathbf{x}$ of the training set to form the trees. The quality of the branch split is measured using the mean squared error (MSE) of the regression: $ {\rm MSE} = \frac{1}{N} \sum_{n=1}^{N} (y_n - \hat{y}_n)^2$.

For each testing sample $\mathbf{x}$, the prediction is obtained from the averaged prediction of the individual trees: $f(\mathbf{x})$ = $\frac{1}{M} \sum_{=1}^{M}f_m(\mathbf{x})$, where $M$ is the total number of trees and $f_m$ stands for the prediction of each tree model using data $\mathbf{x}$. Random forests are known to be robust against overfitting, and have been widely applied for both regression and classification tasks~\cite{weisberg2005applied}. In addition, random forests offer means of interpreting the model using importance ranking and partial dependence analysis~\cite{Breiman2001,friedman2001greedy}. To train the model for predicting the work function, we `fingerprint' the interface structures in our database with a number of features that are physically meaningful and are expected to be correlated with the work functions. Some of the selected features have be shown previously to be critical to describe phase stability~\cite{thermal_stability}, thermal conductivity~\cite{Mingo}, optical absorption~\cite{Pilania2016,HOIP_bandgap}, superconductivity~\cite{superconductor}, catalytic activity~\cite{PCCP_Montoya}, and fuel-cell performance~\cite{dane_several_feul_cell}. In total 38 features are selected and summarized in Table~\ref{Tab:descriptor}.

\begin{table}[h]
    \centering
    \renewcommand{\arraystretch}{1.5}
    \caption{Atomic descriptors that are selected in this work}
    \begin{tabular}{p{2.1cm} p{5.2cm}}
    Notation        & Definition  \\
    \hline

        $\rm IP_{expt}$, $\rm EA_{expt}$, $\rm IP_{calc}$, $\rm EA_{calc}$          & Experimental and calculated ionization potential and electron affinity\\
        ${\chi}_{\rm P}$                                                                                                    &Pauling electronegativity\\
        $\delta$                                                                                                                &Bonding covalency with oxygen\\
        $r_{s}$, $r_{p}$, $r_{d}$                                                                                      & $s$, $p$ and $d$ valence orbital radii of the element~\cite{Zunger}\\
        $R_{\rm atm}$, $\bar{R}_{\rm ion}$                                                                 & Atomic radii and averaged ionic radii\\
        $\mathscr{P}$                                                                                                        & Pettifor's chemical scale~\cite{PETTIFOR}\\
        $Z$                                                                                                                          &Atomic number\\   
        $\mathscr{M}$                                                                                                       &Mendeleev number\\
        $\bar E_{p}$                                                                                                           &$p$ band center in bulk perovskite\\
        $\theta_{d}$, $\theta_{e_{\rm g}}$, $\theta_{t_{\rm 2g}}$                                  &Filling factor of $d$ band, $e_{\rm g}$ and $t_{\rm 2g}$ in bulk perovskite\\
        $\bar E_{2p}^{\rm O}$                                                         			          &Center of oxygen 2$p$ band in bulk perovskite\\
        ${\chi}_{\rm M}^{\rm ABO_3}$                                                           		&Geometric mean of the electronegativity of the perovskite constituents on a Mulliken scale \\
  
    \end{tabular}
    \label{Tab:descriptor}
\end{table}

All features, except $\bar E_{2p}^{\rm O}$ and ${\chi}_{\rm M}^{\rm ABO_3}$, are selected for both A and B elements. We computed the ionization potential $\rm IP_{\rm calc}$ and the electron affinity $\rm EA_{\rm calc}$ of the atoms using the energy of the half occupied Kohn-Sham orbital~\cite{PRL_Bigdata}. Furthermore, the band center of orbital $\varphi$ is the energy difference between the weighted center of the $\varphi$-projected band and the Fermi level in a crystal:

\begin{equation}
        \bar E_{\varphi} = \frac{\int\limits_{-\infty}^\infty E \rho_{\varphi}(E) dE }{\int\limits_{-\infty}^\infty \rho_{\varphi}(E) dE} -E_{\rm F},
\end{equation}
and the filling factor of the $\varphi$ band (similarly for $e_{\rm g}$ and $t_{\rm 2g}$ bands) is calculated from

\begin{equation}
        \theta_\varphi = \frac{\int\limits_{-\infty}^{E_{\rm F}} \rho_\varphi(E) dE}{\int\limits_{-\infty}^\infty \rho_{\varphi}(E) dE},
\end{equation}
where $\rho_\varphi$ stands for the projected density of states of the $\varphi$ orbital. This projection is expressed as

\begin{equation}
        \rho_\varphi(E) = \frac 1{N_{{\boldsymbol k} }} \sum_{n{\boldsymbol k}, \sigma } \int |\langle \psi_{ n {\boldsymbol k} }^{\sigma} | \varphi \rangle |^2 \delta( E - E_{ n {\boldsymbol k}}^{ \sigma } ) dE,
\end{equation}
where $n$, $\boldsymbol k$ and $\sigma$ denote the band index, ${\bf k}$-points and the spin states of the wave function $\psi$, respectively.

We note that the inclusion of DFT features requires some initial bulk calculations. Constructing models using only readily available features such as elemental properties \cite{Magpie, Magpie_glass, Morgan_ML} would overcome this requirement; however, these DFT features enable us to establish closer correlation between the work functions and electronic-structure properties, as further analyzed and discussed below.

\section{Results and discussion}
\label{sec:Results and discussion}
 
\subsection{Random forest regression}
\label{sec:Random-forest-regression}

We develop the random forest models using the {\sc scikit-learn} library~\cite{scikit}. The dataset contains 1248 interface work functions and is composed of an equal amount of AO and BO$_2$ surfaces. Two models are trained independently on the work functions of the AO and BO$_2$ terminations. Before training the models, we note that some of the selected features are correlated. Although such correlations would not impact the performance of the model, they could deteriorate its interpretability. This is because the correlated features carry similar information, thus the feature importance would be shared among them, causing a `dilution' of the importance score across the feature group. Therefore, we carried out a reduction of the feature dimension using the Pearson correlation analysis, as detailed in Supplementary Figure~S1. This process reduces the number of features from 38 to 21 by eliminating the most highly correlated ones. We start the analysis by using all the features to train the random forest regression models. For both AO and BO$_2$ terminations, we partition the dataset into $80\%$ and $20\%$ for training and test set. Using the training set, the hyperparameters that gives the lowest root mean squared error (RMSE) are selected. The RMSE is evaluated with fivefold cross-validations. The obtained model's performance is then validated using only the test set. Such evaluation is repeated 40 times by shuffling the datasets to obtain an averaged total performance. By doing so, we can consistently evaluate the accuracy of the model.

\begin{figure}
 	\centering
 	\includegraphics[width=0.48 \textwidth]{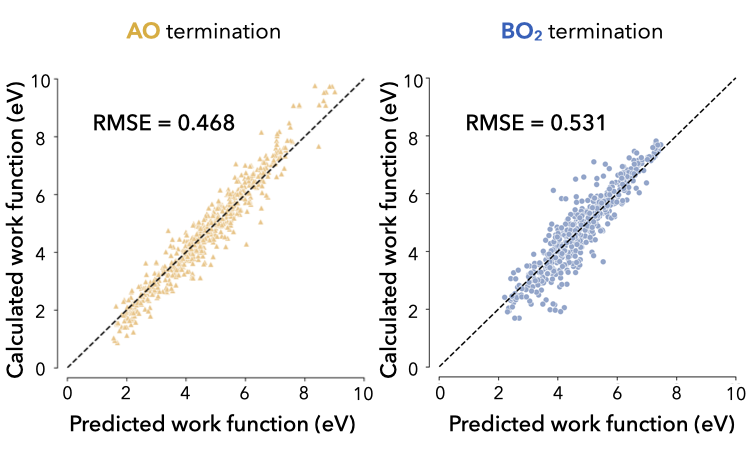}
 	\caption{Predicted versus computed work functions of AO and BO$_2$ terminations. The performance of the random forest regression models is evaluated by averaging the results from randomly shuffled datasets. The averaged root mean squared error for AO and BO$_2$ terminations are 0.468 eV and 0.531 eV, with standard deviations of 0.047 and 0.048, respectively.}
 	\label{random_forest}
\end{figure}

\begin{figure}
 	\includegraphics[width=0.48\textwidth]{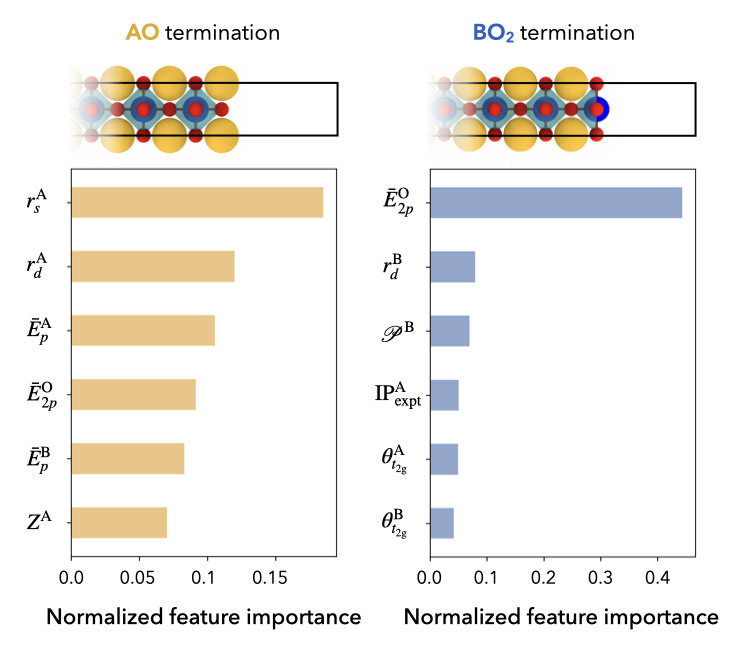}
 	\caption{Normalized feature importance of the 6 most relevant features for AO and BO$_2$ terminations.}
 	\label{importance_ranking}
\end{figure}

We first aim to identify the features that are relevant to the work functions. This is achieved by examining their importance score. In specific terms, the importance of a feature measures how much the feature would impact the predictions. For example, we can calculate the importance of a feature by adding up the weighted variance reduction for all nodes that use this feature as the splitting feature, and then averaged over the trees in the trained forests. Based on the importance score, we perform recursive feature elimination~\cite{superconductor,HOIP_bandgap}, and then re-train the model each time to obtain a new importance ranking. To optimize the model's performance when each feature is removed, the hyperparameter is re-selected using aforementioned process. The yielded model contains compatible hyperparameter and number of features. The averaged RMSE of the regression with respect to the number of features are shown in Supplementary Figures~S2, along with detailed descriptions of the model constructing process. We find that the averaged RMSE of the work function of AO and BO$_2$ terminations are 0.468 eV and 0.531 eV, respectively. The averaged predicted work function $\Phi$ are plotted against the DFT values in Fig.~\ref{random_forest} (a) and (b). The prediction accuracy is reasonable considering that the work functions span a range of 9 eV. For the six most predominant features that are identified in the models, we summarize their normalized feature importance in Fig.~\ref{importance_ranking}. 

Based on the importance ranking, we find that the most relevant features for both surfaces show a consistent pattern despite the different surface structures. For the BO$_2$ terminations, the work function is strongly influenced by the the bulk 2$p$ band center of oxygen $\bar E_{2p}^{\rm O}$, which has an importance score of 0.44. Following that, two features that are related to the terminated element, namely $r_{d}^{\rm B}$ and $\mathscr{P}^{\rm B}$, are found to be relatively important for the work functions of BO$_2$ terminations. This indicates that the work function of BO$_2$ is largely determined by its bulk properties. On the other hand, though $\bar E_{2p}^{\rm O}$ is still relevant to the AO work functions, it only ranked as the fourth most important feature. The first three features are consistently correlated to the surface species of AO terminations; they are $r_{s}^{\rm A}$, $r_{d}^{\rm B}$, and $\bar E_{p}^{\rm A}$. This result shows that, in contrast to BO$_2$ surfaces, the surface contribution to the work function is more significant than the bulk properties for the AO terminations~\cite{dane_WF}. In general, the machine-learning model correctly recognizes the termination effect, where the valence orbital radii for A and B elements are predicted to be among the most essential features for the AO and BO$_2$ terminations. In addition, it is known that the work function is influenced by both bulk and surface properties~\cite{dane_WF,prb_wf,HALAS2010486}. The machine-learning models correctly capture this dependence.

\begin{figure*}[th!]
 	\centering
 	\includegraphics[width=0.96\textwidth]{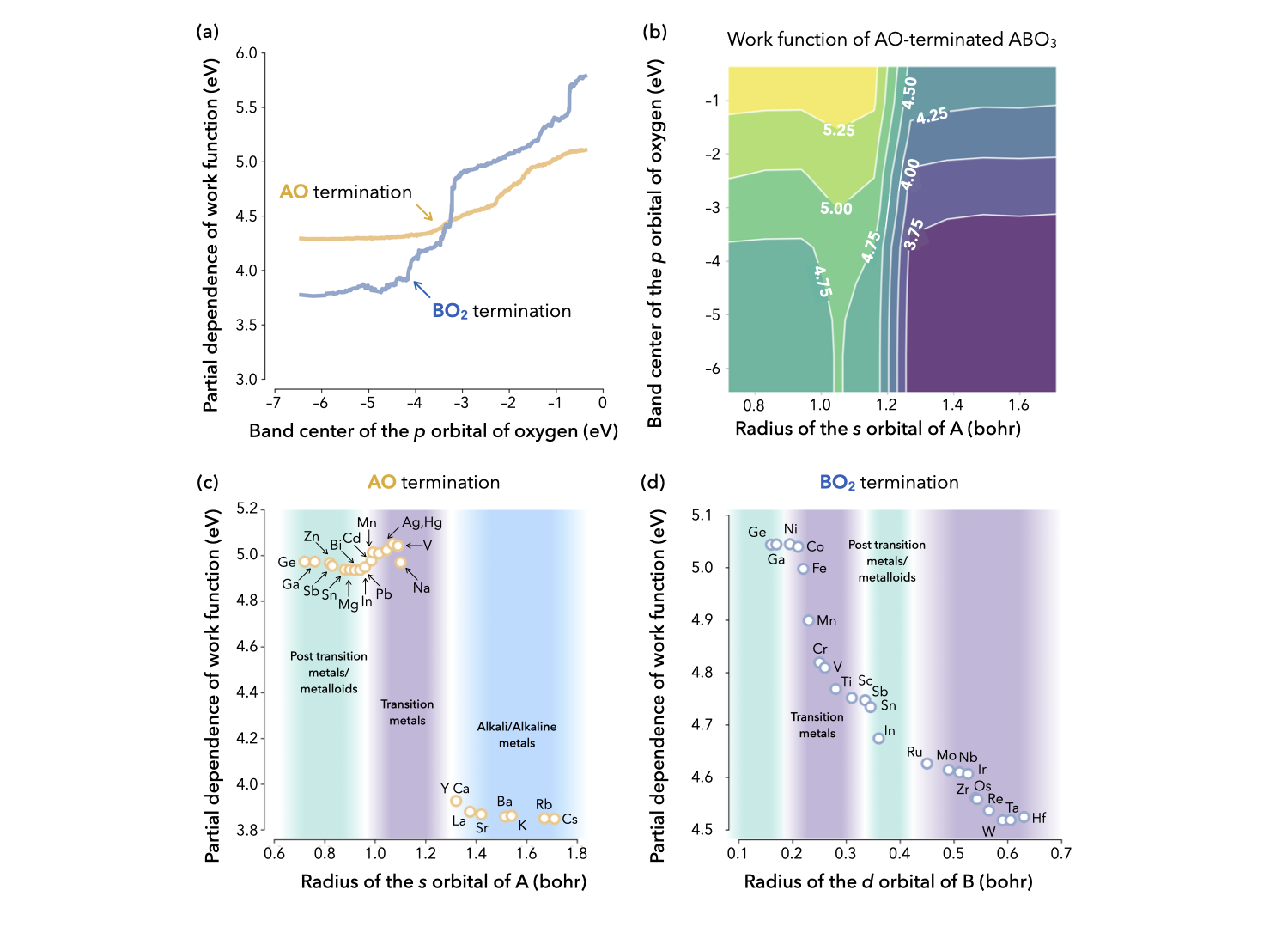}
 	\caption{Partial dependence plots for the AO and BO$_2$ terminations. (a) Dependence of the work function with respect to the orbital energy $\bar E_{2p}^{\rm O}$ for both interfaces. (b) Two-variable dependence plots of the work functions of both orbital energy $\bar E_{2p}^{\rm O}$ and the orbital radius $r_{s}^{\rm A}$ for AO-terminated interfaces. (c, d) Partial dependence of the work function with respect the radius of the $p$ ($d$) orbital of the A (B) element for the AO (BO$_2$) termination. The colored regions represent groups of elements in the periodic table.}
 	\label{PDP}
\end{figure*}

It is interesting to note that, for both AO and BO$_2$ terminations, the energy of the oxygen 2$p$ orbital in the bulk phase plays a critical role. In fact, $\bar E_{2p}^{\rm O}$ is an important bulk electronic predictor that has been used to describe many electronic properties of perovskites, including vacancy formation energies~\cite{MAYESHIBA201671}, oxygen reduction reactivity of oxide fuel cell~\cite{dane_several_feul_cell} and oxygen evolution reactivity~\cite{C5CP04248H,dane_CM}. Specifically for work functions, Jacobs~\textit{et al.} have reported $\bar E_{2p}^{\rm O}$ as a critical descriptor by exploring 20 technologically relevant perovskite materials that are composed of Sr and La for the A atoms and 3$d$ transition metals for the B atoms~\cite{dane_WF}. Here our data-driven approach corroborates that $\bar E_{2p}^{\rm O}$ remains an effective descriptor even for a wide range of metastable perovskites.

Although it helps identify the most significant features, the importance score only indicates how much the predictions are affected by the features, without explaining the specific relationship. To answer this question, we conduct a partial dependence analysis~\cite{friedman2001greedy} for the two most predominant features. Partial dependence plots (PDP) illustrate the marginal effect (in the probabilistic sense) of the selected features on the predictions after integrating out the other variables. If we only focus on one specific feature $\textbf{x}$, the interactions between $\textbf{x}$ and the response of the target can be estimated by marginalizing the predictions over all other features. This partial dependence function $\bar f_s($\textbf{x}$)$ can be expressed as

\begin{equation}
    \bar f_s(\mathbf{x})= \frac 1N \sum_{n=1}^N f(x_{1,n}, \cdots, x_{s-1,n}, x, \cdots,x_{M,n}),
\end{equation}
where $\bar f_s(\mathbf{x})$ is approximated by averaging the output of the trained model for all features except the selected feature $x = x_s$ in the dataset. $M$ is the total number of features in the model, and $N$ is the total number of samples. Similarly to the previous analysis, the PDP is obtained by averaging the results using the shuffled datasets. In Fig.~\ref{PDP}, we illustrate the PDP for both AO and BO$_2$ work functions with respect to $\bar{\rm O}_{2p}$ and the valence orbital radii $r_{s}^{\rm A}$ and $r_d^{\rm B}$.

\begin{figure*}[h!]
 	\centering
 	\includegraphics[width=0.90\textwidth]{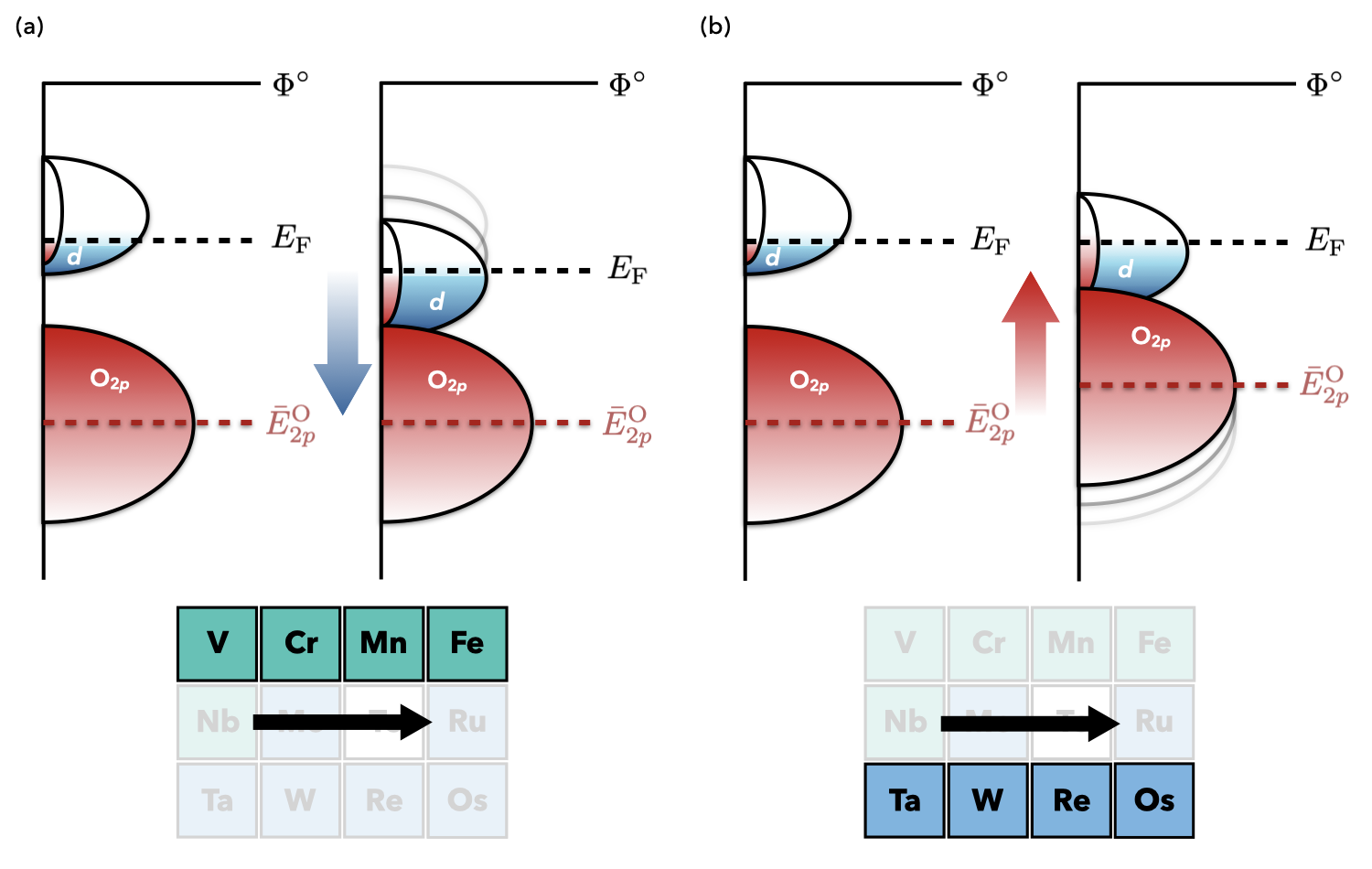}
 	\caption{Schematized densities of states for perovskites that contain (a) 3$d$ transition metals and (b) 5$d$ transition metals. $\Phi^\circ$ stands for the energy level of vacuum. The red and blue regions correspond to the $p$ and $d$ bands of oxygen and of the B-site transition metal, respectively. The black and red dash lines represent the Fermi level and the O 2$p$ band center.}
 	\label{band_diagram}
\end{figure*}

Despite different surface structures, we find that the general trend of how $\bar E_{2p}^{\rm O}$ influences $\Phi$ is universal, as shown in Fig.~\ref{PDP} (a): with a larger separation between $\bar E_{2p}^{\rm O}$ and the Fermi level (more negative band center of O 2$p$ in bulk), the work function shows an approximately monotonic decrease. Interestingly, such correlation starts to break down for the perovskite interfaces with low work functions, where the work function reaches a plateau when the $\bar E_{2p}^{\rm O}$ is below --4 eV, especially for AO termination. To explain these trends, we examine the density of states, and the correlations between $\bar E_{2p}^{\rm O}$ and the work function. In general, the low work function of a perovskite originates from low filling of the $d$ bands, as shown in Fig.~\ref{band_diagram}(a). One of the representative compounds of this class is SrVO$_{3}$~\cite{dane_WF}. Thus, as we move across the 3$d$ transition metal series, the $d$ bands are filled up with electrons and move down to hybridize with the O ${2p}$ band. This can also be understood by analyzing electron affinities. A more electronegative B site will create a more covalent bond with oxygen, thus leading to more pronounced band hybridization. A key characteristic for such hybridization is that, the band center of oxygen 2$p$ is almost unchanged with respect to vacuum level [see Supplementary Fig.~S5(a)]. This observation is consistent with previous literature~\cite{dane_WF} and enables one to understand the linear correlation between the $\bar E_{2p}^{\rm O}$ and the work functions: with increasing $d$ filling, the d bands hybridize with the O 2$p$ bands and reduce the energy separation between the Fermi level and the O $2p$ band center. Since $\bar E_{2p}^{\rm O}$ remains almost constant with respect to the vacuum level (indicating moderate charge transfer between the inner and outer layers), a decrease in the Fermi level and an increase in the work function is observed.

Yet, we observe that the previously described correlation breaks down for deep $\bar E_{2p}^{\rm O}$ levels.  To understand these deviations, we examined the compounds with $\bar E_{2p}^{\rm O}$ deeper than --4 eV and found that those perovskites primarily contain 5$d$ elements, such as Ta, W and Re. By examining their projected density of states [see Supplementary Fig.~S5(b)], we found that the key difference lies in the stability of the band center for O 2$p$.  In this case, it is observed that $\bar E_{2p}^{\rm O}$ is no longer constant (indicating charge transfer between the inner and outer layers), as depicted in Fig.~\ref{band_diagram}(b) by the shift of $\bar E_{2p}^{\rm O}$ towards the vacuum energy level. This trends explains the loss of correlation between the $\bar E_{2p}^{\rm O}$ level and the work function of those compounds.  We further study this trend by examining the partial dependence of the work function with respect to the $\bar E_{2p}^{\rm O}$ and $r_{s}^{\rm A}$ for the AO termination in Fig.~\ref{PDP}(b). It is apparent that when $\bar E_{2p}^{\rm O}$ is above --4 eV, the isocontours align horizontally, which indeed confirms the strong correlation between $\bar E_{2p}^{\rm O}$ and the work functions. In contrast, the isocontours are mostly vertical when $\bar E_{2p}^{\rm O}$ is deeper than --4 eV.

Next, we turn our attention to the influence of the valence orbital radii on the work function. We first discuss the AO termination, and we highlight different groups of elements in Fig.~\ref{PDP} (c). We find that the work function can be parsed into three regions: (1) $r_{sp}^{\rm A}$ $<$ 1.0 bohr, (2) 1.0  bohr $<$ $r_{s}^{\rm A}$ $<$ 1.3  bohr, and (3) $r_{s}^{\rm A}$ $>$1.3 bohr. In fact, these three regions correspond to alkali/alkaline-earth metals, transition metals, and post-transition metals and metalloids. The valence orbital radii have been shown to capture the periodic trends~\cite{Zunger}, except for Li and Na due to their small radii. For elements belonging to the family of post-transition metals and metalloids, the work function tends to decrease with respect to the increase of the valence orbital radii. This can be explained in terms of the electronegativity: alkali/alkaline-earth metals (larger $r_{s}^{\rm A}$) show lower electronegativity compared to that of post-transition metals and metalloids (smaller $r_{s}^{\rm A}$), thereby yielding smaller work functions. In addition, because of our choice of A cation across the periodic table, we observe a clear separation of the work function between surfaces that are terminated with alkali/alkaline-earth metals and the post-transition metals/metalloids. The low work function of the alkali/alkaline terminated perovskites make them potential candidates for designing thermionic converters.

We can now discuss the trend between the size of the $d$ orbital radii of the B atoms ($r_d^{\rm B}$) and the work function of BO$_2$ surfaces with similar arguments. Figure~\ref{PDP}(d) shows that the work functions also decrease with $r_d^{\rm B}$. The increasing $r_d^{\rm B}$ radii reflect a decrease in electronegativity, thus causing a diminution in the work function. Compared to the PDP of the AO surface, we do not observe significant separation in the dependence of the work functions as a function of BO$_2$ surface interactions. This is likely due to the fact that the B cations are mainly composed of the transition metals and metalloids, with no alkaline/alkali metals included. These partial dependence analyses reveal that the work function is determined by both bulk electronic properties and surface electronegativity. By controlling compositions and structures, these two effects can be leveraged simultaneously to design materials with desired work functions. 

In closing, we underscore the practical importance of our statistical observations. A low work function is a crucial requirement for designing electron emitters and thermionic energy converters, and we find here that perovskites with alkali or alkaline-earth metals at the A site are promising candidates for these applications, as shown in Fig.~\ref{PDP}. We conclude that although the center of the oxygen 2$p$ band is a sensitive descriptor of the work function for a number of perovskites, AO-terminated surfaces with shallow Fermi energy are much better described by the orbital radii of the A-site elements. This analysis demonstrates the possibility of optimizing surface structure and chemistry to effectively reduce the work function for {\it e.g.} thermionic energy conversion.

\section{Conclusions}
We have examined the work functions of cubic perovskites by statistical means. We have constructed a database of perovskites and have employed a random forest regression to predict their work functions, achieving predictive accuracy with only a few features included. Two central features that primarily control the perovskite work functions have been identified: the oxygen 2$p$ band center and the valence orbital radii of the surface-terminating cations. The oxygen 2$p$ band center is found to be crucial to the determination of the BO$_2$-termination work functions, while $r_{s}^{\rm A}$ predominantly influences the AO-termination work functions. We have explained how those electronic descriptors affect the work functions of perovskites using partial dependence analysis, and have found that the general trends are related to the stability of oxygen energy levels and atomic electronegativities. These correlations may benefit the search for metal oxides with desired surface electronic properties. For instance, optimizing the compositions of the perovskites to achieve deep oxygen 2$p$ band centers while simultaneously terminating the interface with alkali or alkaline-earth elements may yield optimally low work functions, which is essential for thermionics. Conversely, the perovskites that have shallow oxygen 2$p$ band centers, coupled with $p$-block metal or metalloid terminations, may be of interest for designing hole collectors.

\section*{Conflicts of interest}
The authors declare no competing conflicts of interest.

\section*{Acknowledgements}
Y.~X.~and I.~D.~acknowledge primary financial support from the National Science Foundation under grant DMREF-1729338. W.~C. acknowledges support from the National Science Foundation under grant DMR-2011839 through the MRSEC Center for Nanoscale Science of the Pennsylvania State University. The computational work was performed using the Roar supercomputer of the Penn State Institute for Computational and Data Sciences.

\section*{Data availability }

The work-function and descriptor databases, and the code repository are available online at https://github.com/yyx5048/Cubic-Perovskite-WorkFunction.



\balance


\bibliographystyle{rsc} 
\providecommand*{\mcitethebibliography}{\thebibliography}
\csname @ifundefined\endcsname{endmcitethebibliography}
{\let\endmcitethebibliography\endthebibliography}{}

\end{document}